\documentstyle[12pt,epsf,rotate]{article}
\input FEYNMAN
\newcommand{\pr}{\paragraph{}}
\newcommand{\nn}{\nonumber}
\newcommand{\be}{\begin{equation}}
\newcommand{\ee}{\end{equation}}
\newcommand{\bea}{\begin{eqnarray}}
\newcommand{\eea}{\end{eqnarray}}

\begin{document}
\newcommand{\nd}[1]{/\hspace{-0.5em} #1}
\begin{titlepage}

\begin{flushright}
NTUA 67/97 \\
OUTP-97-58P \\
cond-mat/9710288 \\
\end{flushright}

\begin{centering}
\vspace{.05in}
{\Large {\bf Gauge-theory approach to planar doped antiferromagnets 
and external magnetic fields$^{\dagger}$ \\}}
 
\vspace{.1in}
 K. Farakos 
 
\vspace{.05in}

National Technical University of Athens, 
Physics Department,
Zografou Campus
GR-157 73, Athens, Greece, \\

\vspace{.05in}
and \\
\vspace{.05in}
N.E. Mavromatos$^{*}$ \\
\vspace{.05in}
University of Oxford, Department of (Theoretical) Physics, 
1 Keble Road OX1 3NP, Oxford, U.K. \\

\vspace{.2in}
{\bf Abstract} \\
\vspace{.05in}
\end{centering}
{\small A review is given of a relativistic non-Abelian gauge theory 
approach to the physics of spin-charge separation 
in doped quantum antiferromagnetic planar systems,  
proposed recently 
by the authors. 
Emphasis is put on the effects of constant
external magnetic fields on excitations about the 
superconducting state in the model. 
The electrically-charged Dirac fermions (holons), 
describing excitations about specific points on the fermi surface, 
e.g. those corresponding to the nodes of a $d$-wave superconducting gap 
in high-$T_c$ cuprates, 
condense, resulting in 
the opening of a Kosterlitz-Thouless-like gap (KT) at such nodes. 
In the presence of strong external
magnetic fields at the surface 
regions of the planar superconductor, in the direction 
perpendicular to the 
superconducting planes,  
these KT gaps appear to be enhanced. 
Our preliminary analysis, based on 
analytic Scwhinger-Dyson treatments, 
seems to indicate that for an {\it even} number of Dirac fermion 
species, required in our model as a result of  
gauging a particle-hole $SU(2)$ symmetry, Parity or Time Reversal violation 
does not necessarily occurs.
Based on these considerations,  
we argue that recent experimental findings, concerning 
thermal conductivity plateaux of quasiparticles in planar high-$T_c$ cuprates
in strong external magnetic fields, 
may indicate the presence of such KT gaps, caused by  
charged Dirac-fermion excitations
in these materials, as suggested in the above model.}

\vspace{.4in}

\begin{flushleft} 
October 1997 \\
$^{\dagger}$ Based on Invited talk by N.E.M. at the 
`5th Chia Workshop on Common Trends in Particle and Condensed 
Matter Physics', 
Conference Center, Grand-Hotel 
Chia-Laguna, Chia (Sardegna), Italy, 1-11 September 1997. \\
$^{(*)}$~P.P.A.R.C. Advanced Fellow. \\
\end{flushleft} 

\end{titlepage}

\section{Introduction}
\pr
Gauge symmetry breaking without an elementary Higgs particle,
which proceeds via the dynamical formation of fermion condensates,  
has been a fascinating idea, which however had been tried
rather inconclusively, so far, in attempts to understand 
either chiral 
symmetry breaking in four-dimensional QCD via a new strong 
interaction (technicolour)~\cite{xsb}, or in 
the breaking of a local gauge symmetry through the formation 
of pair condensates in non-singlet channels~\cite{dimop}.
In all such scenaria the basic idea is that there exists an 
energy scale at which the gauge coupling becomes strong enough 
so as to favour the formation of non-zero fermion condensates
$<{\bar f} f'>$ which are not invariant under the 
global or local symmetry in question. 
It is the purpose of this talk to point out that 
similar scenaria of dynamical gauge symmetry breaking 
in three-dimensional gauge theories~\cite{farak} lead to interesting 
and unconventional superconducting properties of the theory
after coupling to electromagnetism~\cite{dor},  
and therefore may be 
of interest to condensed matter community, especially 
in connection with the high-temperature superconductors.

In a recent publication~\cite{fm} we 
have argued that the doped 
large-$U$ Hubbard (antiferromagnetic) models possess
a {\it hidden} local {\it non-Abelian} $SU(2)\times U_S(1)$ 
phase symmetry related to spin interactions.
This symmetry 
was discovered using an appropriate `particle-hole symmetric
formalism' for the electron operators~\cite{zou}, 
and employing a generalised 
{\it slave-fermion} ansatz for {\it spin-charge 
separation}~\cite{Anderson}, 
which allows intersublattice 
hopping for holons, and hence spin flip~\footnote{Non-abelian 
gauge symmetry structures 
for doped antiferromagnets,
in a formally different context though, i.e. by employing slave-boson
techniques,
have also been proposed by other authors~\cite{leewen}. 
However, the patterns of symmetry breaking 
discussed here, and in ref. \cite{fm},  
are physically different from those 
approaches, and they allow for a unified description 
of slave-boson and slave-fermion approaches 
to spin-charge separation.}. 
The spin-charge separation 
may be physically interpreted as implying an effective 
`substructure' of the electrons due to the 
many body interactions in the medium. 
This sort of idea, originating from Anderson's 
RVB theory of spinons and holons~\cite{Anderson},
was also pursued recently by 
Laughlin, although from a (formally at least)
different perspective~\cite{quark}. 

The effective long wavelength 
model of such a statistical system 
is remarkably similar to a 
three-dimensional gauge model 
of particle physics proposed in ref. \cite{farak}
as a toy example for chiral symmetry breaking 
in QCD. In that work, it has been argued that {\it dynamical 
generation} of a fermion mass gap 
due to the 
$U_S(1)$ subgroup of $ SU(2) \times U_S(1)$
breaks the $SU(2)$ subgroup down to a $\sigma_3-U(1)$ 
group, where $\sigma_3$ is the $2\times 2$  Pauli matrix. 
{}From the particle-theory view point 
this is a Higgs mechanism without an {\it elementary} 
Higgs excitation. 
The analysis carries over to the condensed-matter case,
if one associates the mass gap to the holon condensate~\cite{fm}. 
The resulting effective theory of the light degrees of freedom 
is then similar to the 
continuum limit of \cite{dor}
describing unconventional parity-conserving superconductivity.  
Parity conservation is a result of the existence of an {\it even} 
number of fermion species, due to the underlying $SU(2)$ 
particle-hole spin symmetry~\cite{fm,dor}. Energetics in such systems
prevent the dynamical generation of a parity violating gap~\cite{Vafa,app}.  

The effect of external magnetic fields on the state of the condensate
is of extreme interest, in view of recent experiments
with high-$T_c$ cuprates pertaining to the 
thermal conductivity of quasiparticle (QP) excitations 
in the superconducting state~\cite{ong}. QP thermal conductivity
plateaux in a high-magnetic-field phase of these materials 
indicate~\cite{ong} the opening of a gap for strong magnetic fields,
depending on the intensity of the external field, when the latter
is applied perpendicularly
to the cuprate planes. 
As we shall argue in this talk, the phenomenon 
appears to be generic 
for systems of charged Dirac fermions in external magnetic fields,
in the sense that strong enough magnetic fields are capable of 
inducing spontaneous formation of neutral condensates, 
whose magnitude scales with the magnetic field
strength~\cite{miransky,hong,ng}. Such gaps disappear at critical temperatures
proportional to the size of the gap. Despite the time-reversal 
breaking by the magnetic field source, 
there is no evidence in such systems
for parity violating gaps if there is an {\it even number} 
of fermionic flavours, as is the case of the model of ref. \cite{fm}. 
However this issue is still not quite settled, 
and deserves further non-perturbative
studies, for reasons that we shall discuss at the end of the talk. 
In this latter respect we should mention the recent interpretation  
of refs. \cite{haldane,laughlinT} about the experimental findings of ref. 
\cite{ong}. According to those scenaria, 
the high-magnetic field phase of the cuprates,
is characterized by a transition of the 
superconducting state to a {\it parity- and time-reversal- broken} state, 
proposed by Laughlin 
some time ago~\cite{anyon}. In view of the results presented here, however, 
we shall argue that this appears not to be necessarily the case 
in the presence of
an {\it even number of fermion species}, as in the model 
of ref. \cite{fm,dor}. This would allow a smooth connection 
of the high-magnetic-field phase gap with the zero-field 
superconducting gap of the model of ref. \cite{dor,fm}.

It is important to notice that 
the experimenal indication on the opening of a gap 
in ref. \cite{ong} at the nodes of the $d$-wave superconducting gap,
at least in the high-magnetic field phase, will be argued 
to imply an experimental test of the model proposed in ref. \cite{dor} and 
\cite{fm}, involving Dirac fermions to describe excitations 
about {\it specific points}, such as the gap nodes, 
of the fermi surface of doped $t-j$ models. This is due to the fact that
Dirac fermion condensation can be triggered by strong magnetic 
fields~\cite{miransky}, in a Scwhinger-Dyson treatment of the
relativistic  
field theory proposed in ref. \cite{fm}, and in ref. \cite{dor}.
Such a gap  
reduces to the superconducting gap of ref. \cite{dor} 
in the limit of vanishing external fields, provided that the 
gap-inducing statistical gauge interactions (due to the spin-spin 
interactions of the holons) is strong enough. A 
rather preliminary account 
of these considerations will be given below. 
More detailed investigations 
will be the subject of forthcoming work~\cite{future}. 
The important point in the
opening of this gap is that due to the planar character of the 
holon excitations in these model, the opening of a gap, 
in the absence of a magnetic field, occurs in the Kosterlitz-Thouless
mode~\cite{RK,dor}, i.e. occurs without implying 
the existence of a local order parameter,
as a result of strong phase fluctuations. 
Hence the $d$-wave character of the superconducting state is
preserved at the nodes, 
despite the opening of the gap there.
Unfortunately at present, as far as we understand, 
the experiments cannot reach the 
weak (vanishing) magnetic field region, which could be a good testing 
ground for the above ideas. 
 
\section{Low-Energy Limit of Doped Planar Antiferromagnets 
about specific points on the Fermi Surface}

We would like 
to start our discussion
by 
considering  
the low-energy limit 
of {\it doped antiferromagnetic} planar systems
with specific points on their fermi surface (e.g. nodes, or 
points where a d-wave gap vanishes, etc). Such systems might be 
of
relevance to the physics of high-temperature
superconductors, since recently it is believed that 
high-temperature superconductivity in cuprates 
is highly anisotropic and the gap symmetry is $d$-wave~\cite{dwave},
with the gap vanishing along lines of {\it nodes} 
on the Fermi surface.

We shall be very brief in our discussion here.
For more details we refer the reader to ref. \cite{fm}
and references therein. 
The model considered in \cite{fm} was the strong-U Hubbard model,
describing doped antiferromegnets with the constraint 
of no more than one elelctron per lattice site.
The key suggestion in ref. \cite{fm}, which lead to 
the non-abelian gauge symmetry structure for the doped antiferromagnet,
was the {\it slave-fermion} spin-charge separation ansatz 
for physical electron operators 
at {\it each lattice site} $i$~\cite{fm}:
\be
\chi _{\alpha\beta,i} 
\equiv \left(
\begin{array}{cc}
c_1 \qquad c_2 \\
c_2^\dagger \qquad -c_1^\dagger \end{array}
\right)_i \equiv {\widehat \psi} _{\alpha\gamma,i}{\widehat z}_{\gamma\beta,i} =
\left(\begin{array}{cc}
\psi_1 \qquad \psi_2 \\
-\psi_2^\dagger \qquad \psi_1^\dagger \end{array}
\right)_i~\left(\begin{array}{cc} z_1 \qquad -{\overline z}_2 \\
z_2 \qquad {\overline z}_1 \end{array} \right)_i 
\label{ansatz2}
\ee
where 
$c_\alpha$, $\alpha=1,2$ are electron anihilation
operators, the Grassmann variables $\psi_i$, $i=1,2$ 
play the r\^ole of holon excitations, while the bosonic
fields $z_i, i=1,2,$ represent magnon excitations~\cite{Anderson}.
The ansatz (\ref{ansatz2}) 
has spin-electric-charge separation, since only the 
fields $\psi_i$ carry {\it electric} charge.
This ansatz characterizes the proposal 
of ref. \cite{fm} for the dynamics underlying 
{\it doped} antiferromagnets.
In this context, the 
holon fields ${\widehat \psi} _{\alpha\beta}$ 
may be  
viewed as substructures of the 
physical electron $\chi_{\alpha\beta}$~\cite{quark},
in close analogy to the `quarks' of $QCD$.   

As argued in ref. \cite{fm} 
the ansatz is characterised by   
the following {\it local}
phase (gauge) symmetry structure: 
\be
  G=SU(2)\times U_S(1) \times U_E(1) 
\label{group}
\ee
The 
local 
SU(2) symmetry is discovered if one defines the transformation 
properties of the ${\widehat z}_{\alpha\beta}$ 
and ${\widehat \psi} ^\dagger_{\alpha\beta}$ 
fields to be given by left multiplication
with the $SU(2)$ matrices, and pertains to the spin degrees of freedom.
The 
local $U_S (1)$ `statistical' phase symmetry, which 
allows fractional statistics of the spin and charge 
excitations. This is an exclusive feature
of the three dimensional geometry, and is similar in spirit
to the bosonization technique of the spin-charge 
separation ansatz of ref. \cite{marchetti},
and allows the alternative possibility 
of representing the holes as slave bosons and  
the spin excitations as fermions. 
Finally the $U_E(1)$ symmetry is due to the electric
charge of the holons. 

The pertinent long-wavelength gauge model, describing the low-energy 
dynamics
of the large-U Hubbard 
antiferromagnet, in the spin-charge separation phase (\ref{ansatz2}),
assumes the form~\cite{fm}: 
\bea
&~&H_{HF}=\sum_{<ij>} tr\left[(8/J)\Delta^\dagger_{ij}\Delta_{ji}
+ K(-t_{ij}(1 + \sigma_3)+\Delta_{ij}){\widehat \psi}_j V_{ji}U_{ji}
{\widehat \psi}_i^\dagger\right] + \nn \\ 
&~&\sum_{<ij>}tr\left[ K{\overline {\widehat z}}_iV_{ij}U_{ij}{\widehat z}_j\right] + h.c. 
\label{Hub}
\eea
where $J$ is the Heisenberg antiferromagnetic interaction, 
$K$ is a normalization constant, and 
$\Delta_{ij}$ is a Hubbard-Stratonovich field that linearizes
four-electron interaction terms in the original Hubbard model, 
and 
$U_{ij}$,$V_{ij}$ are the link variables for the $U_S(1)$ and 
$SU(2)$ groups respectively. 
The conventional lattice gauge theory form of the action (\ref{Hub}) 
is derived upon freezing the fluctuations of the $\Delta_{ij}$ 
field~\cite{fm}, and   
integrating out the magnon fields, $z$,
in the path integral. This latter operation yields 
appropriate Maxwell kinetic terms for the link variables 
$V_{ij}$, $U_{ij}$, 
in a low-energy derivative expansion
~\cite{IanA,Polyakov}.
On 
the lattice such kinetic terms 
are given by plaquette terms of the form~\cite{fm}:
\be   
\sum_{p} \left[\beta_{SU(2)}(1-Tr V_p) + \beta_{U_S(1)}(1-Tr U_p)\right]
\label{plaquette} 
\ee
where $p$ denotes sum over 
plaquettes of the lattice, 
and $\beta_{U_S(1)} \equiv \beta_1 $, $\beta_{SU(2)} \equiv \beta_2 = 
4\beta_1$ 
are the dimensionless (in units of the lattice spacing) 
inverse square couplings of the $U_S(1)$ and $SU(2)$ groups,  
respectively~\cite{fm}. The above relation between the $\beta_i$'s
is due to the specific form of the $z$-dependent terms in (\ref{Hub}),
which results in the same induced couplings $g_{SU(2)}^2=g_{U_S(1)}^2$.
Moreover,  there is a non-trivial connection of the 
gauge group couplings to $K$~\cite{fm}: 
\be
K \propto g_{SU(2)}^2 = g_{U_S(1)}^2 \propto J\eta
\label{connection}
\ee 
with $\eta$ the doping concentration in the sample~\cite{fm,dorstat}. 
To cast the 
symmetry structure in a 
form that is familiar to 
particle physicists, one may change representation 
of the $SU(2)$ group, and instead of working with $2 \times 2$ 
matrices in (\ref{ansatz2}), one may use a representation 
in which the fermionic matrices ${\widehat \psi}_{\alpha\beta}$ 
are represented as two-component (Dirac) spinors in `colour' space:
\be
{\tilde \Psi}_{1,i}^\dagger =\left(\psi_1~~-\psi_2^\dagger\right)_i,~~~~
{\tilde \Psi} _{2,i}^\dagger=\left(\psi_2~~\psi_1^\dagger
\right)_i,~~~~~i={\rm Lattice~site} 
\label{twospinors}
\ee

In this representation  
the two-component spinors ${\tilde \Psi} $ (\ref{twospinors}) 
will act as Dirac spinors, and the $\gamma$-matrix (space-time)
structure will be spanned by the irreducible 
$2 \times 2$ representation. 
By 
assuming a background $U_S(1)$ 
field of flux $\pi$ per lattice plaquette~\cite{dor},
and considering quantum fluctuations around this background
for the $U_S(1)$ gauge field, 
one can show that there is a Dirac-like structure 
in the fermion spectrum describing the excitation 
about a node in the fermi surface~\cite{Burk,AM,dor,dorstat}. 
This leads to a conventional 
Lattice gauge theory form for the effective low-enenrgy Hamilonian of the
large-$U$,  doped Hubbard model~\cite{fm}. Remarkably, 
this lattice gauge theory has the 
same form as (\ref{effeaction}). The constant $K$ of (\ref{effeaction}) can then be identified
with $K$ in (\ref{Hub}).  

In the above context, a strongly coupled
$U_S(1)$ group can dynamically generate a mass gap 
in the holon spectrum~\cite{app,dor,kocic,koutsoumbas,maris}, 
which breaks the $SU(2)$ local symmetry
down to its Abelian subgroup generated by 
the $\sigma_3$ matrix~\cite{farak,fm}.  
{}From the view point of the statistical model (\ref{Hub}), 
the breaking of the $SU(2)$ symmetry down to its Abelian 
$\sigma_3$ subgroup may be interpreted as  
restricting the holon hopping effectively to 
a single sublattice, since 
the intrasublattice hopping is suppressed 
by the mass of the gauge bosons.  
In a low-energy
effective theory of the massless degrees of freedom 
this reproduces the 
results of ref. \cite{dor,Sha}, derived 
under a large-spin 
approximation for the antiferromagnet, $S \rightarrow \infty$, 
which is not necessary 
in the present approach. 

The (naive continuum limit of) low-energy theory 
about such nodes of the fermi surface of the 
planar antiferromegnt, then, reads: 
\be 
{\cal L}_2 \equiv =-\frac{1}{4}(F_{\mu\nu})^2 
-\frac{1}{4}({\cal G}_{\mu\nu})^2  +{\overline \Psi}D_\mu\gamma_\mu\Psi
\label{su2action}
\ee
where now $D_\mu = \partial_\mu -ig_1a_\mu^S-ig_2\sigma^aB_{a,\mu}
-\frac{e}{c}A_\mu$,
$B_\mu^a$ is the gauge potential of the local (`spin') $SU(2)$ group, 
and ${\cal G}_{\mu\nu}$ is the corresponding field strength. 
The fermions $\Psi$ are viewed as 
{\it two-component} spinors. 
Once we gauged the $SU(2)$ group, the colour structure is up and above 
the space-time Dirac structure, and in two-component 
notation the $SU(2)$ group is generated by the 
familiar $2 \times 2$ Pauli matrices $\sigma^a$, $a=1,2,3$. 
In this way, the fermion condensate ${\cal A}_3$
can be generated dynamically by means of a strongly-coupled
$U_S(1)$. In this context, energetics 
prohibits the generation of a parity-violating 
gauge invariant $SU(2)$ term~\cite{Vafa}, and so 
a parity-conserving mass term necessarily breaks~\cite{farak}
the $SU(2)$ group down to a $\sigma_3-U(1)$ sector~\cite{dor}, generated
by the $\sigma_3$ Pauli matrix in two-component notation. 
\pr
The above symmetry breaking patterns may be proven analytically~\cite{farak} 
on the lattice, in the strong $U_S(1)$ limit, $\beta_1 \rightarrow 0$.
The 
lattice lagrangian, corresponding to the continuum lagrangian 
(\ref{su2action}),
assumes the form: 
\bea
&~& S=\frac{1}{2}K \sum_{i,\mu}[{\overline \Psi}_i (-\gamma_\mu) 
U_{i,\mu}V_{i,\mu} \Psi_{i+\mu}  + \nonumber \\
&~& {\overline \Psi}_{i+\mu}
(\gamma _\mu)U^\dagger_{i,\mu}V^\dagger_{i,\mu}
\Psi _i ]  \nn \\
&~& + \beta _1 
\sum _{p} (1 - trU_p) + \beta _2 \sum _{p} (1-trV_p) 
\label{effeaction}
\eea
where $\mu =0,1,2$, $U_{i,\mu}=exp(i\theta _{i,\mu})$ 
represents the 
statistical $U_S(1)$ gauge field, $V_{i,\mu}=exp(i\sigma ^a B_a)$ is the $SU(2)$ gauge field, 
The fermions $\Psi $
are taken   
to be two-component (Wilson) spinors, in both Dirac and 
colour spaces~\cite{farak,fm}.
\pr
In the strong-coupling limit, 
$\beta _1 =0$   
the field $U_S(1)$ field
may be integrated out analytically in the path integral
with the result
~\cite{farak}:
\be
Z = \int dV d{\overline \Psi}d \Psi exp(-S_{eff})
\label{action3}
\ee
where
\bea
&~&S_{eff} = \beta _2 \sum _{p} (1 -trV_p)
- \sum _{i,\mu}{\rm ln}I^{tr}_0(2\sqrt{y_{i\mu}}) \nn \\
&~& y_{i\mu} = -\frac{K^2}{4}tr[M^{(i)}(-\gamma _\mu)V_{i\mu}
M^{(i+\mu)}(\gamma_\mu)V_{i\mu}^\dagger] 
\label{action4}
\eea
where 
and 
\be
  M^{(i)}_{ab,\alpha\beta} 
\equiv \Psi _{i,b,\beta}{\overline \Psi}_{i,a,\alpha},~~a,b={\rm
colour},~\alpha,\beta={\rm Dirac},~i={\rm lattice~site}
\label{mesons}
\ee
are the meson states, and 
${\rm ln}I^{tr}_0$ denotes the logarithm of 
the zeroth order Modified Bessel function~\cite{Abra}, 
{\it truncated }
to an order determined by the number of the Grassmann (fermionic) 
degrees of freedom in the problem~\cite{kawamoto}. 
In our case, due to the $SU(2)$ and spin quantum numbers of the {\it 
lattice spinors} $\Psi$, one should retain terms in $-{\rm ln}I^{tr}_0$  
up to ${\cal O}(y^4)$:
\be
-{\rm ln}I^{tr}_0(2\sqrt{y_{i\mu}})=
-y_{i\mu}+\frac{1}{4}y^2_{i\mu}
-\frac{1}{9}y_{i\mu}^3 +\frac{11}{192}y_{i\mu}^4 
\label{trunclog}
\ee
The above expression 
is an {\it exact} result, irrespectively of the magnitude 
of $y_{i\mu}$. 
\pr
The low-energy (long-wavelength)
effective action is written as a path-integral 
in terms of gauge and meson fields,
$
Z=\int [dV dM]exp(-S_{eff}+ \sum_{i}tr{\rm ln}M^{(i)})$, 
where the meson-dependent term
comes from the Jacobian in passing from 
fermion integrals to meson ones~\cite{kawamoto}.
To identify the symmetry-breaking patterns of the 
gauge theory (\ref{trunclog})
one may concentrate on 
the lowest-order terms in $y_{i\mu}$, 
which will yield the gauge boson 
masses. Higher-order terms will describe interactions. 
Symmetry-breaking patterns 
for $SU(2)$ will emerge out of a non zero VEV 
for the meson matrices $M^{(i)}$. This is confirmed by a 
detailed analysis presented in refs.~\cite{farak,fm}, where we refer
the interested reader.

\begin{centering}
\begin{figure}[htb]
\epsfxsize=2in
\centerline{\epsffile{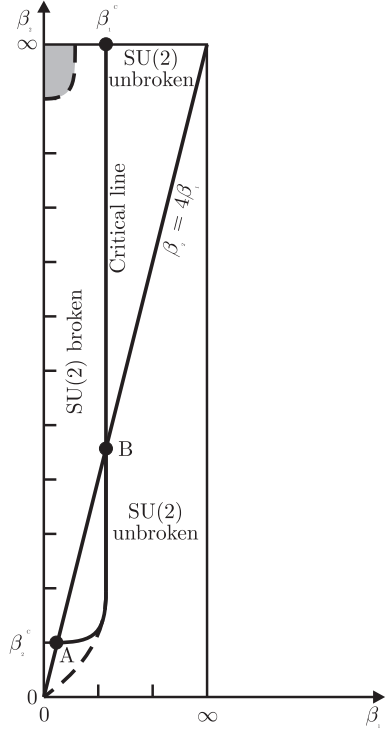}}
\vspace{1cm}
\caption{{\it Phase diagram 
for the $SU(2) \times U_S(1)$ gauge theory.
The critical line separates the phase
of (dynamically) broken $SU(2)$ symmetry 
from the unbroken phase. Its shape is conjectural 
at this stage, in particular with respect to 
the order of  magnitude
of $\beta_2^c$. The shaded region has been 
analysed in ref. \cite{farak}. 
The dashed line 
represents a probable critical line  
in the case of the statistical model of ref. \cite{fm}.
The straight line indicates the specific relation 
of the coupling constants in that model.}} 
\label{fig1}
\end{figure}
\end{centering}

In ref. \cite{fm} a phase diagram was conjectured, which is depicted in 
fig. 1. The diagram indicates the existence of a `critical line' in 
coupling constant space, seprating the phases of broken from unbroken 
$SU(2)$ gauge symmetry. In gauge models describing low-energy effective 
theories of 
realistic systems, 
the various couplings depend on the doping concentration in the 
sample~\cite{dor}, and so various 
regions of the phase diagram 
are reached by varying the doping concentration~\cite{fm}.
It is of interest for what follows to concentrate briefly on the 
line $\beta_2=0$ of the phase diagram. The existence of a critical 
$U_S(1)$ coupling above which dynamical mass generation for fermion 
occurs has been confirmed both in Scwhinger-Dyson (SD) 
(large flavour number $N$)~\cite{app,dor,maris} 
and lattice~\cite{kocic,koutsoumbas} treatments. For instance, in large-$N$
SD treatments, the dynamically generated fermion gap in a theory 
with an even number of fermion flavours $N$, is parity conserving, 
and is given by:
\be 
   m_{s} =\alpha 
{\rm exp}\left(-\frac{2\pi}{\sqrt{\frac{32}{\pi^2 N} - 1}}\right)
\label{flavour}
\ee
where $\alpha \equiv g_s^2/4\pi$, with $g_s$ the dimensionful coupling 
(dimensions 
of $\sqrt{{\rm mass}}$) 
of a (2+1)-dimensional gauge theory and  
$N$ is the number of (four-component) fermion flavours. For dynamical 
mass to occur $N$ must be less than $32/\pi^2$.   
For such a $N =O(1)$ (\ref{flavour}) implies that $m_s << \alpha$,
and thus dynamical mass generation is an infrared (low-energy) phenomenon. 
For instance for $N=1$ (one four component spinor) one gets
$m_{s} \simeq 0.015\alpha$. 

The above analysis treated the flavour number as a fixed number. 
According to ref. \cite{amelino}, however, 
the non-trivial infrared dynamics of gauge theories in three dimensions
induces a Wilsonian-type renormalization-group slow `running' 
of the effective flavour number with the momentum scale. This behaviour is  
obtained by integrating out
field modes in quantum loop diagrams 
with momenta below an infrared cut-off. Such a procedure,
when combined with an 
appropriate 
analysis of the renormalization-improved Schwinger-Dyson 
equations, yields an `asymptotically-free', slow running, $N(p/\alpha)$,
whose increase towards low momentum 
scales is cut-off at a finite value $N^*$, at 
the low momentum scales $p << \alpha $ 
appropriate for 
dynamical mass generation. 
It is $N^*$ that 
should enter the gap formula (\ref{flavour}), which thus may lead to a 
significant increase in the gap magnitude, since $N ^*$ is {\it smaller}
than the 
bare $N$, which is reached only~\cite{amelino} for ultraviolet momentum scales
$p >> \alpha $. These considerations 
should be born in mind when one makes attempts
to connect the above results with realistic
situations concerning high-temperature superconductors. 
For instance, the above-mentioned non-trivial infrared structure 
may be responsible for a non-fermi liquid behaviour of the materials 
in their normal phase, 
where no dynamical mass generation occurs~\cite{amelino}. 

At finite temperatures the gap disappears at a critical temperature
$T_c$ which is proportional to the size of the gap at zero temperature~\cite{dor}:
\be 
     k_B T_c \simeq \frac{2}{n} m_s(T=0)
\label{gapT}
\ee
where $k_B$ is Boltzmann's constant, and 
the number $n$ is of order ${\cal O}(6)-{\cal O}(10)$.  
The ambiguity in this number is due to the approximations 
employed in his computation. In ref. \cite{dor}, where only 
the instantaneous approximation for the statistical SD photon 
propagator has been considered, $n ={\cal O}(10)$. However,
going beyond the instantaneous approximation reduced this number 
to about ${\cal O}(6)$~\cite{aitchT}.

\section{Superconducting Properties}

As the next topic of our generic analysis 
of three-dimensional gauge models we would like to discuss 
the superconducting consequences of 
the above dynamical breaking patterns of the $SU(2)$ group. 
{\it Superconductivity} is obtained upon coupling 
the system to external elelctromagnetic potentials, 
which leads to the presence of an additional gauge-symmetry, 
$U_E(1)$, that of ordinary electromagnetism.

\begin{centering}
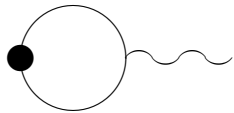
\begin{figure}[htb]
\vspace{2cm}
%
%\begin{figure}[h]
\bigphotons
%\label{vacpol3}
\begin{picture}(30000,5000)(0,0)
\put(20000,0){\circle{100000}}
%\drawline\gluon[\E\REG](22000,0)[4]
\drawline\photon[\E\REG](22000,0)[4]
\put(18000,0){\circle*{1000}}
\end{picture}
%\end{figure}
%\centerline{\epsffile{fig2.eps}}
%
\vspace{1cm}
\caption{{\it Anomalous one-loop Feynman matrix element,
leading to a Kosterlitz-Thouless-like breaking of the 
electromagnetic $U_{E}(1)$ symmetry, and thus 
superconductivity, once a fermion 
mass gap opens up. The wavy line represents the $SU(2)$ 
gauge boson $B_\mu^3$,
which remains massless, while the blob denotes an insertion 
of the fermion-number
current  $J_\mu={\overline \Psi}\gamma_\mu \Psi$.
Continuous lines represent fermions.}}
\label{fig2}
\end{figure}
\end{centering}

Upon the 
opening of a mass gap in the fermion (hole) spectrum, one obtains 
a non-trivial result for the 
following
Feynman matrix element: 
${\cal S}^a = <B^a_\mu|J_\nu|0>,~a=1,2,3$, with $J_\mu ={\overline
\Psi}\gamma _\mu \Psi $, the fermion-number current.
Due  
to the colour-group structure, only the massless $B^3_\mu $ 
gauge boson of the $SU(2)$ group, corresponding to the $\sigma _3$
generator in two-component notation, contributes to the 
matrix element. 
The non-trivial result for the matrix element ${\cal S}^3$
arises from an {\it anomalous one-loop graph}, depicted in 
figure 2, and it
is given by~\cite{RK,dor}:
\be
    {\cal S}^3 = <B^3_\mu|J_\nu|0>=({\rm sgn}{M})\epsilon_{\mu\nu\rho}
\frac{p_\rho}{\sqrt{p_0}} 
\label{matrix2}
\ee
where $M$ is the parity-conserving fermion mass 
(holon condensate), generated dynamically 
by the $U_S(1)$ group. As with the 
other Adler-Bell-Jackiw anomalous graphs in gauge theories, 
the one-loop result (\ref{matrix2}) 
is {\it exact} and receives no contributions from higher loops~\cite{RK}. 
\pr
This unconventional 
symmetry breaking (\ref{matrix2}), 
does {\it not have a local order parameter}~\cite{RK,dor},
since the latter is inflicted by strong phase fluctuations, 
thereby resembling  the
Kosterlitz-Thouless
mode of symmetry breaking. The {\it massless} Gauge Boson 
$B_\mu^3$ of the 
unbroken $\sigma_3-U(1)$ subgroup of $SU(2)$ is responsible for the 
appearance of a {\it massless pole} in the electric current-current 
correlator~\cite{dor}, which is the characteristic feature 
of any {\it superconducting theory}. As discussed in ref. \cite{dor},
all the standard properties of a superconductor, such as 
the Meissner
effect, infinite conductivity, flux quantization, London action etc. are 
recovered in such a case. 
The field $B^3_\mu$, or rather its {\it dual} $\phi$ defined by
$\partial _\mu \phi \equiv \epsilon_{\mu\nu\rho}\partial_\nu B^3_\rho$,
can be identified with the Goldstone
Boson of the broken $U_{em}(1)$ (electromagnetic) symmetry~\cite{dor}. 

It is important to notice that 
the absense of a local order parameter of the above mechanism 
implies that, upon interpreting the phenomenon as being associated
with the opening of a Kosterlitz-Thouless 
gap at the {\it nodes} of the original $d$-wave 
superconducting gap of the cuprate, 
the opening of such gap does not affect the $d$-wave
nature of the state. 

\section{Effects of an external magnetic field} 

In this section we shall discuss the effects of an external 
magnetic field in charged excitations about the superconducting state.
Due to the Meissner effect, the bulk of the superconductor 
is shielded off from  external magnetic fields. However, in the surface region
the magnetic field penetrates the superconductor. It is in such regions
that the discussion below refers to. 
\pr
The formalism we shall follow is essentially due to Schwinger~\cite{schwing},
who was actually the first to compute exactly the fermion 
propagator in the presence of a constant external magnetic field. 
This formalism has been applied recently to discuss 
the effect of external fields in inducing fermion 
condensates~\cite{miransky,hong,ng}, the latter
being defined as the coincidence limit of the configuration 
space Dirac propagator in the presence of the external field.  

The result of 
such analyses was that, in the context of gauge theories, like quantum
electrodynamics in three and four dimensions, strong external 
magnetic fields
are capable of inducing fermion condensates proportional to some 
power of the external field intensity. At finite temperatures
the condensate disappears at a critical temperature proportional to the 
induced gap at zero temperatures. 
In particular, for four-dimensional QED, the gap was 
proportional to the square root of the magnetic field strength, and 
consequently the transition temperature. Such features, 
especially the transition temperature dependence 
on the gap, seem to 
characterize the experiment of ref. \cite{ong} on the behaviour of the 
thermal conductivity in the presence of magnetic fields. 

What we shall argue in this section is that, if we apply 
the above formalism to the model of ref. \cite{fm}, a similar 
dependence on the magnetic field strength in the high-field phase
is obtained for the induced dynamical gap. What is important 
to notice is that in that model the gap seems to be enhanced 
by the presence of the strong magnetic field, given that the 
same set of Schwinger-Dyson equations, used to study the dynamical opening 
of a gap in the absence of magnetic fields in the superconducting phase
of the model of ref. \cite{fm,dor}, is also used in the presence of background 
external fields in the surface regions of the superconductor (where 
the magnetic field penetrates the sample), and the limit to the 
vanishing field phase seems to be obtained smoothly, at least formally.
There are some questions regarding the existence of a critical magnetic 
field which might induce a transition, but we shall discuss this point 
later, as at present we cannot perform complete analytic computations
in the model of ref. \cite{fm,dor}. 

We start our discussion by presenting a general discussion on the 
induction of a Dirac fermion condensate in three-dimensional systems 
in the presence of magnetic fields. The magnetic field can always 
be considered as truly four dimensional~\cite{dor}, but the charged Dirac
fermions (holons) will be considered as genuinely three dimensional. 
When we discuss dynamical opening of a mass gap, the three-dimensional 
Schwinger-Dyson equations will be obtained by naive dimensional 
reduction of four-dimensional ones.

Let one consider a system of an even number of Dirac fermions
(like the one considered in the previous section (\ref{su2action}) 
in the presence of an external magnetic field $B$. 
For the time being 
we assume that we are already in the superconducting phase of the 
model, due to the statistical $U_S(1)$ interactions, which implies 
that the fermions acquire a parity conserving mass $+ m,-m$ 
in two-component notation. Our point is to examine the 
effect of the externally applied magnetic field. This means that 
from now on we may ignore the statistical interactions in the 
model (\ref{su2action}), and replace the effective action with that of 
two species of (2+1)-dimensional free Dirac 
fermions $\Psi_{1,2}$ in the presence of an external 
gauge potential 
\be
A_\mu^{ext} = -Bx_2\delta_{\mu1}
\label{extpot}
\ee 
The Lagrangian is:
\be
  L=\frac{1}{2}{\overline \Psi}(i \gamma^\mu (\partial_\mu - ieA_\mu^{ext}) - m)\Psi  
\label{lagrangian}
\ee
where $m$ is a {\it parity conserving } bare fermion 
mass, and the $\gamma$-matrices
belong to the reducible  $4\times 4$ representation, appropriate 
for an even number of fermion species formalism~\cite{app,dor,fm}. 

This problem has been studied in ref. \cite{miransky}, 
and below we sketch the derivation for the benefit of the 
non-expert readers. 
The induced fermion condensate is given by the coincidence 
limit of the fermion propagator
\bea 
 &~&  <0|{\overline \Psi} \Psi |0> = -Lim_{x \rightarrow y}trS(x,y) \nn \\
 &~& {\rm where}~~ S(x,y)=<0|T{\overline \Psi (x)}\Psi (y)|0>
 \label{propagator}
\eea

Following the proper time formalism of Schwinger~\cite{schwing},
the propagator $S(x,y)$ 
in the presence of a constant external magnetic field 
can be calculated exactly~\cite{miransky}:  
\bea
&~&S(x,y) = {\rm exp}\left(ie\int_x^y A^{ext}_\mu dz^\mu \right)
{\tilde S}(x-y) \\
&~&{\tilde S}(x) =\int_0^\infty \frac{ds}{8(\pi s)^{3/2}}{\rm exp}[-i
\left(\frac{\pi}{4} + sm^2 \right)]\times {\rm exp}[-\frac{i}{4s}
x_\nu C^{\nu\mu}x_\mu] \nn \\
&~&\times [\left(m + \frac{1}{2s}\gamma^\mu C_{\mu\nu}x^\nu-\frac{e}{2}\gamma^\mu
F_{\mu\nu}^{ext}x^\nu \right) \times \left(esB{\rm cot}(eBs)-\frac{es}{2}
\gamma^\mu\gamma^\nu F_{\mu\nu}^{ext} \right)]
\label{proptime}
\eea
where $F_{\mu\nu}^{ext}$ is the Maxwell tensor corresponding to the 
external background gauge potential (\ref{extpot}), and 
$C_{\mu\nu}=\eta^{\mu\nu}+ ([F^{ext}]^2)^{\mu\nu}[1-eBs{\rm cot}(eBs)]/B^2$,
and the line integral is calculated  along a straight line. 
A useful expression, to be used in the following, is the Fourier 
transform in Euclidean space 
of ${\tilde S}(x-y)$, 
${\tilde S}_E(k)$~\cite{miransky}:
\bea
&~&{\tilde S}_E (k) =-i\int_0^\infty ds {\rm exp}[-s(m^2 +k_0^2 + {\underline k}^2
\frac{tanh(eBs)}{eBs}] \nn \\
&~&\times \{[-k_\mu\gamma^\mu + m -i(k_2\gamma_1 -
k_1\gamma_2)tanh(eBs)]\times [1 -i\gamma_1\gamma_2tanh(eBs)]\}
\label{eucls}
\eea
A straightforward calculation, then, yields in (2+1)-dimensional
space times the following result for the 
magnetic-field induced condensate in the limit where the bare mass
$m \rightarrow 0$~\cite{miransky}: 
\bea
&~&<0|{\overline \Psi} \Psi |0>=-\frac{i}{(2\pi)^3}\int d^3k tr {\tilde S}_E(k)
\nn \\
&~&=-Lim_{\Lambda \rightarrow \infty} Lim_{m \rightarrow 0} 
\frac{4m}{(2\pi)^3}
\int d^3k \int_{1/\Lambda^2}^\infty ds {\rm exp}[-s(m^2 + k_0^2 +{\underline k}^2 \frac{tanh(eBs)}{eBs})] \nn \\
&~&=-Lim_{\Lambda \rightarrow \infty} Lim_{m \rightarrow 0} \frac{m}{2\pi^{3/2}}
[\pi^{1/2}|eB|\frac{1}{m} + {\cal O}(\frac{1}{\Lambda})] =-\frac{|eB|}{2\pi}
\label{condensate}
\eea
thereby implying that in three dmensional gauge theories 
a strong magnetic field (such that $|eB| $ is much larger 
than any other mass scale in the problem), may induce a fermion gap. 
Notice an important difference from the corresponding four-dimensional 
problem, where the corresponding fermion condensate reads: 
\be 
  <{\overline \Psi}\Psi > \sim -|eB|\frac{m}{4\pi^2} (
{\rm ln}\frac{\Lambda ^2}{m^2} + {\cal O}(m^0) ) \rightarrow 0,
\qquad ; m \rightarrow 0
\label{fourdim}
\ee
and tend to zero as the bare mass $m \rightarrow 0$ , 
for a fixed (ultraviolet ) scale $\Lambda$. However, even in four-dimensional 
theories, quantum dynamics of the electromagnetic field may drastically 
change the situation if treated non-perturbatively~\cite{miransky,ng}. 
We shall come back to this point, and the comparison with the 
three dimensional theories, later in the section.

There is an elegant physical interpretation of this phenomenon, which is 
associated with the {\it infrared} physics in the presence of 
strong magnetic fields. According to ref. \cite{miransky}, the problem 
is connected to the energy spectrum of Dirac fermions 
in the presence of strong external fields. For the four component
fermions the energy spectrum is given by the Landau levels:
\bea
&~&   E_0=\pm m \nn \\
&~&E_n = \pm \sqrt{m^2 + 2|eB|n} \qquad n \ge 1.  
\label{landau}
\eea
The density of states with energies $\pm m$ is 
$|eB|/2\pi$, 
whilst for $n \ge 1$ is $|eB|/\pi$, i.e. the condensate
(\ref{condensate}) equals the density of states 
at the lowest Landau level. More precisely, the propagator 
${\tilde S}_E(k)$ can be decomposed over Landau poles~\cite{miransky}:
\bea
&~&{\tilde S}_E (k) =-i{\rm exp}\left(-\frac{{\underline k}_{vertical}^2}{|eB|}\right)
\sum _{n=0}^\infty (-1)^n \frac{D_n}{k_0^2 + m^2 + 2|eB|n} \nn \\
&~&D_n(eB, k)=(m-k_0\gamma_0)
[(1-i\gamma_1\gamma_2{\rm sgn}(eB))
L_n^0(2\frac{{\underline k}_{vertical}^2}{|eB|}) \nn \\
&~&-(1+i\gamma_1\gamma_2{\rm sgn}(eB))
L_{n-1}^0
(2\frac{{\underline k}_{vertical}^2}{|eB|})] \nn \\
&~&+4(k_1\gamma_1 + k_2\gamma_2)
L_{n-1}^1(2\frac{{\underline k}_{vertical}^2}{|eB|}) 
\label{poles}
\eea
where ${\underline k}_{vertical} $ denotes momentum components in a direction 
perpendicular to that of the applied field, and 
$L_n^a$ are Laguerre polynomials, with $L_{-n}^a=0, n >0$. 
In the limit $m << |eB|$, the condensate appears due to the 
lowest Landau level: 
\be
 <0|{\overline \Psi}\Psi |0> \simeq -\frac{m}{2\pi^3} \int 
d^3k \frac{{\rm exp}(-{\underline k}_{vertical}^2/|eB|)}{k_0^2 + m^2}=-\frac{|eB|}{2\pi}
\label{lowestlevelcond}
\ee

The above consideration can be made more precise to the 
more realistic case of the model (\ref{su2action}), where one 
examines the effects of a strong external magnetic field 
on the dynamical mass generation for fermions due to the statistical 
$U_S(1)$ interaction. 
In four dimensional Abelian gauge theories, due to (\ref{fourdim}),
the quantum dynamics of the Abelian gauge field is the dominant one.
However, as mentioned above in three dimensions this may not be the case,
since free three-dimensional fermions in an external field 
may condense for strong enough fields (\ref{condensate}).
In the three-dimensional case, of the model of ref. \cite{fm} then, 
the r\^ole of the bare mass $m$ in (\ref{condensate}) 
may be played by the dynamically-generated mass due to the 
statistical $U_S(1)$ interactions, in which case the holons
are driven to a new enhanced gap 
\be 
    m_B^{cl} = \sqrt{|<{\overline \Psi} \Psi >|} \sim \sqrt{|eB|}
\label{holongap}
\ee
under the influence of a strong magnetic field. 
We stress again that the above relation (\ref{holongap})
is striclty valid in {\it three} dimensional gauge theories.
\pr
A finite-temperature $T$ (Matsubara) analysis 
can be performed 
by compactifying the time direction 
in (\ref{lowestlevelcond}), which
results in a discrete spectrum for the energies $k_0 \rightarrow 
\omega_n = 2\pi (n + \frac{1}{2})T$.
In the small mass $m << \sqrt{eB}$ regime, 
the finite-temperature condensate 
becomes: 
\be
<0|{\overline \Psi}\Psi |0>_T \simeq -\frac{m T|eB|}{\pi} 
\sum_{n=-\infty }^{\infty}  
\frac{1}{4\pi^2T^2(n+\frac{1}{2})^2 + m^2} =
-\frac{|eB|}{2\pi}{\rm tanh}\frac{m}{2T}
\label{lowestlevelcondT}
\ee
thereby implying the {\it absence} of the condensate 
for 
{\it any finite tmperature} if 
the bare infrared cut-off mass $m \rightarrow 0$. 
\pr
What we shall do next is to examine the 
effects of the quantum dynamics of the $U_S(1)$ field in the 
model (\ref{su2action}) 
on the gap (\ref{holongap}). 
We shall argue, following 
analyses~\cite{miransky,ng}, that 
such quantum effects are capable of generating dynamically a small 
$m << \sqrt{eB}$, which then leads to an enahnaced gap 
(\ref{lowestlevelcondT}), under the influence of a strong 
(external) magnetic field. In this approach, the critical temperature 
coincides with the critical temperature at which the quantum 
corrections $m$ disappear. 

In order to  get an estimate of 
the quantum corrections, and their dependence on the magnetic field
intensity, 
we shall follow a 
rather cavalier approach and, 
instead of 
solving a three-dimensional Schwinger-Dyson equation from first principles
(see figure 3),
we shall use results from corresponding 
treatments in four dimensions~\cite{miransky}, and dimensionally reduce them.

\begin{centering} 
\begin{figure}[htb]
\bigphotons
\label{SDvacpol3}
\begin{picture}(40000,10000)(0,-3000)
\drawline\fermion[\E\REG](5000,0)[5000]
\put(\pmidx,\pmidy){\circle*{1000}}
\global\advance\pfrontx by -500
\global\advance\pfronty by -200
\put(\pfrontx,\pfronty){$($}
\global\advance\pbackx by 300
\put(\pbackx,\pfronty){$)^{-1}$}
\global\advance\pbackx by 2000
\put(\pbackx,-200){=}
\global\advance\pbackx by 2000
\drawline\fermion[\E\REG](\pbackx,0)[5000]
\global\advance\pfrontx by -500
\global\advance\pfronty by -200
\put(\pfrontx,\pfronty){$($}
\global\advance\fermionbackx by 300
\put(\fermionbackx,\pfronty){$)^{-1}$}
\global\advance\fermionbackx by 2000
\put(\fermionbackx,-200){$-$}
\global\advance\fermionbackx by 2000
\drawloop\gluon[\NE 3](\fermionbackx,\pbacky)
\global\advance\fermionbackx by 2300
\global\advance\pbacky by 1900
\put(\fermionbackx,\pbacky){\circle*{1100}}
\global\advance\pbacky by -1900
\put(\pbackx,\pbacky){\circle*{1000}}
\global\advance\pbackx by 1250
\drawline\fermion[\W \REG](\pbackx,\pbacky)[6700]
\global\advance\pbackx by 3000
\put(\pbackx,\pmidy){\circle*{1000}}
\end{picture}
\vspace{1cm}
\caption{{\it The Schwinger-Dyson equation for the fermion self-energy.
The curly line indicates the $U_S(1)$ statistical photon. 
Straight lines denote fermions in the presence of the external 
magnetic field. Blobs indicate quantum corrections (loops), 
which are ignored 
in the ladder approximation. 
Quantum dynamics of the 
electromagnetic field has been suppressed.}} 
\end{figure}
\end{centering}

In three dimensions the 
r\^ole of the quantum fluctuations of the $U_S(1)$ gauge field 
becomes non-trivial due to the 
non-trivial infrared physics~\cite{amelino} (as compared
to the momentum scale $\sqrt{|eB}|$).
For simplicity below we shall concentrate 
on the area of the phase diagram of fig. 1 where the $SU(2)$ 
coupling constant vanishes, $\beta_2 \rightarrow \infty$, since its presence
will not alter qualitatively the results.  
Dynamical mass generation in the presence of an external field,
due to quantum dynamics for the gauge field,  
has been studied in four-dimensional models in refs. \cite{miransky,ng}. 
For our condensed-matter related problem we shall assume that the 
electromagnetic field is genuinly four-dimensional projected on the 
Cu-O planes~\cite{dor}. This implies that the three-dimensional gauge theory
(\ref{su2action}) may be viewed as being obtained by 
appropriate dimensional reduction of a four-dimensional theory. 
We shall implement dimensional reduction at the level of the 
non-perturbative gap equations that describe dynamical mass generation,
in order to study the effects of the quantum 
dynamics of the statistical gauge field $U_S(1)$ 
on the mass gap (\ref{holongap}).

There are two equivalent formalisms one can follow in studies of 
dynamical mass generation, that of Bethe-Salpeter equation~\cite{miransky},
and that of Scwhinger-Dyson equations~\cite{ng}. Both make use of the 
expression (\ref{poles}) for the fermion propagator in 
external fields. The statistical $U_S(1)$ gauge interactions in the model 
(\ref{su2action}) can be treated as usual, either in large-N expansions~\cite{app,dor} or in the weak coupling regime by summing up ladder graphs. 

In the following we shall be interested in the regime where the 
three-dimensional coupling constant $\alpha = g_s^2/4\pi $ of the $U_S(1)$
interactions in the model (\ref{su2action}), which has dimensions of mass in (2+1)-dimensions,
is much weaker than  
$\sqrt{|eB|}$:
\be  
  \alpha \equiv g_s^2/4\pi << |eB|^{1/2} 
\label{couplingconst}
\ee
Note that in the limit of the external field $B \rightarrow 0$,
a situation which is met in the bulk of a superconductor due to the Meissner 
effect, the dynamical mass generation due to the $U_S(1)$ interactions 
is an infrared phenomenon, which occurs only for coupling constants
relatively strong as compared to the dynamically generated mass~\cite{app}
(c.f. (\ref{flavour})).
What we shall show below is that in the presence of a
strong external magnetic field,
there is an enhancement of this dynamical mass in such a way that the 
mass in the high-field phase, the dynamically generated fermion mass
$m_B$ is much bigger than $m_s$:
\be 
     m_s << \alpha =g_s^2/4\pi << m_B << |eB|^{1/2} 
\label{scales}
\ee
{}From the point of view of a strong field, then, this implies that 
the statistical coupling may be relatively weak, but still dynamical
generation occurs, i.e. the strong magnetic field catalyses 
chiral symmetry breaking~\cite{miransky}. From our 
three-dimensional point of view this mass $m_B$ should be viewed as 
a quantum correction to the magnetically-induced mass gap
$m_B^{cl}$ (\ref{holongap}). Below we shall verify, as a consistency 
check of the approach, that $m_B^{cl} >> m_B$.   

To get a qualitative (preliminary) description of the above phenomenon 
we shall dimensionally reduce the gap equations of 
refs. \cite{miransky,ng}~\footnote{The reader should notice that 
although the methods of the two papers are different, one 
using Bethe-Salpeter~\cite{miransky}, 
the other~\cite{ng,hong},
Schwinger-Dyson treatmnets, however, as should have been expected, the gap 
equation yields similar results.}. The gap equations are derived 
for relatively weak couplings (strong magnetic fields), so that 
only the lowest Landau level pole enters the expression (\ref{poles})
for the fermion propagator ${\tilde S}_E(k)$. Adopting easily the 
situation of ref. \cite{miransky,ng} to our case of the model (\ref{su2action})
one gets, in the strong-field  
limit, the pertinent gap equation
after dimensional reduction to 
(2+1)-dimensions~\footnote{It 
should be mentioned at this point that our dimensional reduction 
occurs up and above the usual reduction $D \rightarrow D-2$ 
which characterizes the physics of 
charged particles in the Lowest landau level~\cite{miransky}. 
Our reduction implies that the fermionic excitations live on a genuine
two-dimensional plane, the Cu-O plane in the model for high-temperature
superconductors~\cite{dor,fm}, and hence any dependence on the 
momentum along the direction of the magnetic field applied perpendicular
to the plane should be ignored.}.

The fact that chiral symmetry breaking 
is catalyzed in that work by strong magnetic fields for 
momentum scales $p \le m_B <<  \sqrt{|eB|}$, implies that 
the ladder approximation for the gauge bosons
may prove sufficient at scales $p \ge \alpha $, 
since the 
$U(1)$ interactions appear relatively weak
at such scales. This is also a valid 
assumption for the model (\ref{su2action}), due to (\ref{scales}). 
In this respect,
it should be pointed 
out that the running of the effective flavour number in three dimensions,
discussed in ref. \cite{amelino}, also supports the 
above point of view, 
due to the asymptotic freedom (with increasing momentum 
scales) of the effective coupling $e_s^2 =N^{-1}(p) \alpha$ 
of the statistical $U_S(1)$ interactions in the model. 
Moreover, the quenched approximation for fermions will also be 
assumed, which from our point of view 
will be sufficient to yield an estimate of the 
corrections to (\ref{holongap}) due to the 
quantum dynamics of the $U_S(1)$ gauge field.

With these in mind, a straightforward analysis, based on that 
of ref. \cite{miransky,ng} extended easily to 
the case of the model (\ref{su2action}), yields the  
following expression for the dimensionally-reduced gap equation:
\be
    1 \simeq \frac{\alpha}{\pi}\int_{-\infty}^{+\infty} \frac{dk}{k^2 + m_{B}^2} \int_0^\infty dx {\rm exp}\left(-\lambda ^2 x/2\right)\frac{1}{k^2 + x}
\label{gap}
\ee
where $\lambda = |eB|^{-1/2}$ is the magnetic length, and $m_B$ is the 
dynamically generated (parity-conserving ) gap in two dimensions, obtained by dimensional reduction of 
the four-dimensional gap equation. 
In the above expression 
$\alpha \propto g_s^2$ is the dimensionful (dimensions of mass) 
fine structure constant
of the statistical $U_S(1)$ interactions of the (2+1)-dimensional 
model (\ref{su2action}), 
which is assumed to satisfy (\ref{scales}). The case $\lambda <<1$ 
can be studied analytically by approximating the gap equation (\ref{gap})
as follows: 
\be
 m_B = 2\alpha  \int_0^\infty dx \frac{e^{-x^2}}
{\frac{m_B}{\sqrt{2|eB|}}+x} 
\sim 2\alpha {\rm ln}[\frac{\sqrt{2|eB|}}{m_B}] \qquad m_B << \sqrt{|eB|} 
\label{gapapprox}
\ee
which results in the following $\sqrt{B}$-dependent gap 
in the high-field phase: 
\be
    m_B \simeq \delta \sqrt{|2eB|}{\rm exp}[-\frac{ m_B}{2\alpha}]
\label{gapfinal}
\ee
where $\delta ={\cal O}(1)$. 
This result is in agreement with our initial assumption (\ref{scales}),
thereby verifying the consistency of the approach. 
A more accurate
analysis involves a numerical evaluation of the integral of the 
right-hand-side of the relation (\ref{gap}).
Defining dimensionless variables $M \equiv m_B/\alpha$,
$K \equiv \sqrt{2|eB|}/\alpha$, one obtains a set of data points 
from 
the numerical evaluation of the integral (\ref{gap}), which then 
is fitted by the approximate function for large $K >> M > 1$:
\be
    m_B/\alpha = C {\rm ln}[\sqrt{2|eB|}/2\alpha]  
\label{fit}
\ee
The data and the fitting curve are presented in 
figure 4.  
{}From this one finds $C \simeq 1.42$.

\begin{centering}
\begin{figure}[htb]
\epsfxsize=4in
\centerline{\rotate{\rotate{\rotate{\epsffile{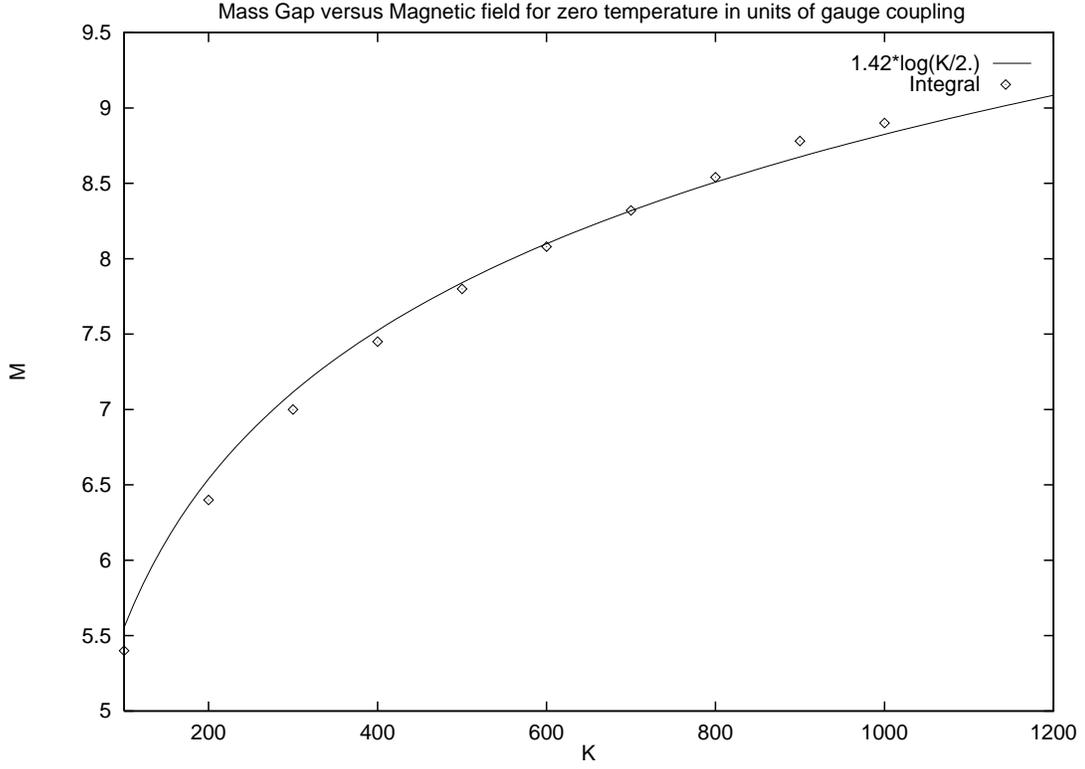}}}}}
\vspace{1cm}
\caption{{\it Graphic solution of (\ref{gap}). Data points  
and fitting curve of the form (\ref{fit}). The fit indicates 
the value $C\simeq 1.42$.
}}
\label{fig3}
\end{figure}
\end{centering}
\pr
It is understood that the analysis above,
should be 
combined, for weak $\alpha$, with the result (\ref{holongap}).
So, {\it qualitatively}, the effects of an external magnetic field 
at zero temperature may be summarized by the enhancement 
of the fermion mass gap in the form: 
\be 
   m_B^{tot} = m_B^{cl} + m_B \sim \sqrt{\frac{|2eB|}{4\pi}} 
+ 1.42 \alpha {\rm ln}[\sqrt{2|eB|}/2\alpha]   
\label{totalgap}
\ee

To get a rough estimate of the critical temperature 
at which the quantum corrections $m_B$,
and according to our previous discussion 
the induced gap itself (\ref{lowestlevelcondT}),  
disappear, 
it suffices to concentrate on the 
finite-temperature formalism 
of the above-described dynamically-generated gap
due to the $U_S(1)$ interactions. 
A finite-temperature $T \ne 0$ (Matsubara) 
analysis of the Schwinger-Dyson equation 
may be performed by dimensional reduction
of the corresponding four-dimensional finite-temperature gap equation
obtained by Gusynin and Shovkovy
in ref. \cite{miransky}. The finite-temperature 
mass gap $m_B(T)$ satisfies the equation (in units where $k_B=1$): 
\be
  1 =2\alpha T\sum_{n=-\infty}^{+\infty} \int_0^\infty dx e^{-\lambda^2 x/2}
\frac{1}{\omega_n^2 + m^2_B(T)}\frac{1}{(\omega_n - \omega_0)^2 + x}
\label{finiteTgap}
\ee
where $\omega_n=2\pi(n + \frac{1}{2})T$. Due to thermal fluctuations 
the induced gap disappears at a critical temperature $T_c$, $m_B(T_c)=0$. 
We are interested in determining a relation between $T_c$ and the externally applied magnetic field.

\begin{centering}
\begin{figure}[htb]
\epsfxsize=4in
\centerline{\rotate{\rotate{\rotate{\epsffile{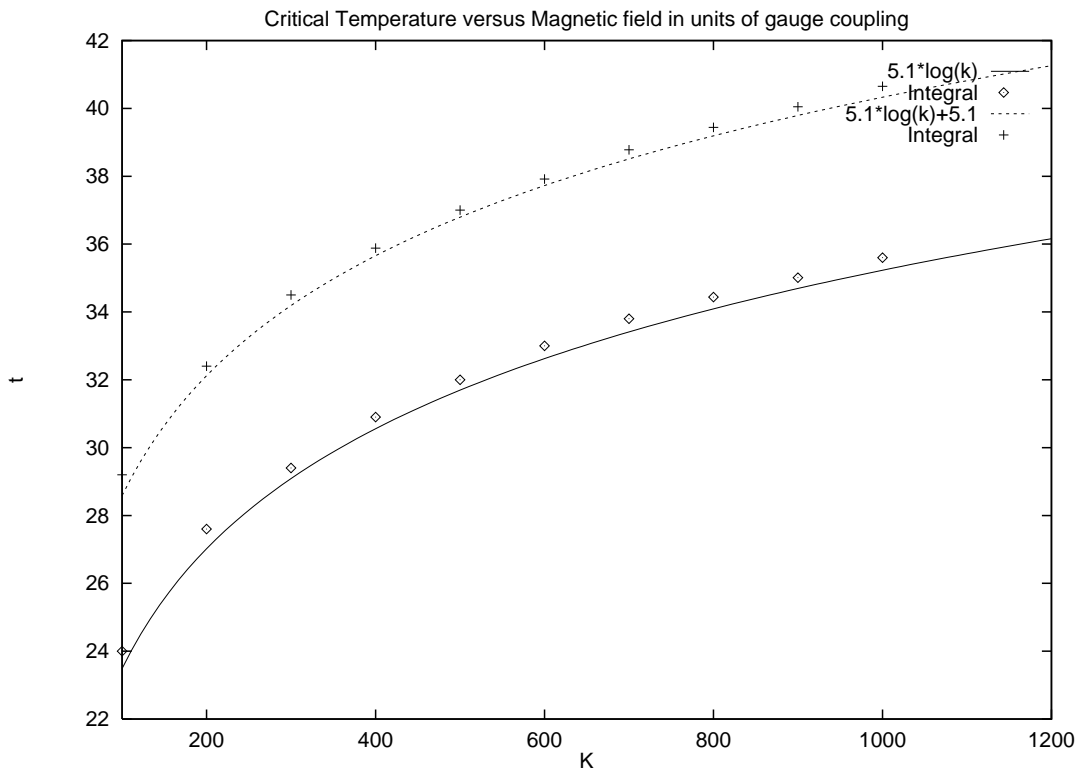}}}}}
\vspace{1cm}
\caption{{\it Numerical evaluation of the ingtegral (\ref{tcgap}), 
which determines $n'$  in 
(\ref{empiricalth2}). Two curves are shown, 
corresponding to  
two different values of the lower limit of the integral. 
For the lower curve, 
corresponding to the value $y_{min} =0.001$ 
for the infrared cut-off, 
the fit indicates 
the average value $n'\simeq 0.65$.
}}
\label{fig4}
\end{figure}
\end{centering}

The critical temperature $T_c$ is determined by the gap equation 
(\ref{finiteTgap}), by setting $m(T_c)=0$ and performing the summation 
over the frequencies. After appropriate rescaling of the
integration variables, the result is:
\be
1 =\frac{\alpha}{2\pi T_c} \int_0^\infty dy {\rm exp}[-\frac{(2\pi T_c)^2}{2|eB|}y]
\frac{1}{(\frac{1}{4} + y)^2}
[\pi (\frac{1}{4} + y) + coth(\pi \sqrt{y})\frac{\frac{1}{4}-y}{\sqrt{y}}]
\label{tcgap}
\ee
Assuming $T_c << \sqrt{2|eB|}$, one observes that 
the integrand is heavily damped for $x > 2|eB|/(2\pi T_c)^2$, so that 
the integration is 
effectively cut-off from above at $x \simeq 2|eB|/(2\pi T_c)^2$. 
The dominant contributions to the integrand of (\ref{tcgap}) 
come from the $coth(\pi\sqrt{y})$-dependent term in the 
infrared regime, $y \rightarrow 0$, and are of logarithmic divergent 
type. A physical way of regulating such divergences 
is by cutting off 
the $\left([{\rm momentum}/2\pi T_c]^2 \equiv y\right)$-integration 
in the infrared region 
by~\footnote{Notice that this infrared cut-off is consistent 
with the ladder approximation for the statistical 
gauge boson $U_S(1)$ in the Schwinger-Dyson equations,
since it implies momenta $k \ge \alpha $,
where, due to the `asymptotic freedom' of the 
running $U_S(1)$ coupling~\cite{amelino}, such an approximation 
proves sufficient, as remarked earlier.
Notice also that this infrared cut-off   
is much less than  
the photon (plasmon) mass at finite temperature which is 
$\sqrt{\alpha}T_c$. Incorporation of the latter 
does not affect the result of the present analysis~\cite{miransky}.} 
$y_{min} = (\alpha/(2\pi T_c))^2 << 1$ we can estimate the result of the 
integration to be:
\be
    \frac{2\pi T_c}{\alpha} \simeq C'{\rm ln}
\left(\frac{\sqrt{2|eB|}}{\alpha}\right)
\qquad ; \qquad C'={\cal O}[1]
\label{empiricalth}
\ee
which is a consistent solution of the mass gap equation under the 
approximations we have made. 
In a more accurate treatment,
the integral (\ref{tcgap}) can be performed 
numerically and is depicted in figure 5, notice that 
$t \equiv 2\pi T_c/\alpha$ and $K \equiv \sqrt{2|eB|}/\alpha$.
For the lower curve, 
corresponding to the value $y_{min} =0.001$ for the   
infrared cut-off,
we can deduce, from figure 5, that $C' \simeq 5.1$. 
Then, from (\ref{fit}),(\ref{empiricalth}), the   
following `empirical' relation emerges (in units where $k_B=1$):
\be
              T_c \simeq n' m_B(T=0) \qquad m_B=E(T=0)/2,~ 
\alpha << \sqrt{|eB|} 
 \label{empiricalth2}
\ee
with $n'\simeq 0.65$, 
and $E(T=0)$ the energy gap at $T=0$.

\section{Conclusions and Outlook: preliminary 
comparison with Experiment}

Some `phenomenology' will help us understand better the 
physics involved in the above phenomenon. 
First we note that, 
due to the relativistic nature of the 
holon excitations 
about the nodes of the d-wave gap~\cite{dor,fm}, 
there is an `effective velocity of light' in the problem, 
which coincides with the 
Fermi velocity for holes $v_F \sim 5 \times 10^{-4} c$. Above we have worked 
in units of $\hbar v_F =1$, so the electric charge, $e$, 
is actually $e/c$, where $e$ is the physical electron charge,
and $c$ is the light velocity. The magnetic length of the problem is therefore 
$(eB/c)^{-1/2}$ in units of $\hbar v_F$ or $e v_FB/\hbar v_F c = eB/\hbar c$
in SI units. This implies that the expressions for the gap
should be multiplied by a factor $\hbar v_F \sim 5 \times 10^{-4}$,
in order to be expressed in $eV$. This will be understood in what follows. 

Let us first remark that for magnetic field of order ${\cal O}(1)-
{\cal O}(10)$ 
Tesla, which is the order of magnitude used in the experiment
of ref. \cite{ong}, and it is below the critical field that destroys
superconductivity in the materials, the classical dynamics of the 
external electromagnetic field seems to indicate 
the formation of a gap (\ref{holongap}) of order: 
\bea 
      m_B^{cl} \sim 0.5 \times  \times 11. 44 \sqrt{|B|/10^4~{\rm Gauss}}~ 
{\rm meV} &\sim & 18~{\rm meV} \qquad B = 10~{\rm Tesla} \nn \\
&\sim & 5.8 ~{\rm meV} \qquad B = 1~{\rm Tesla } 
\label{classical}
\eea

Let us next estimate the size of the 
corrections, $m_B$, due to the quantum dynamics 
of the $U_S(1)$ field. 
In the model 
of ref. \cite{fm} the $U_S(1)$ coupling, $g_s$,  is supposed to
be of the same order as the $SU(2)$ coupling which is 
due to the effective spin-spin interactions of the Heisenberg
antiferromagnetic exchange, $J$. A reasonable estimate will be therefore 
to take the fine structure constant 
of the $U_S(1)$ interactions, $\alpha =g_s^2/4\pi \hbar v_F $ of order 
$\eta J \simeq 0.01 {\rm eV} $,
for a typical (maximum) 
value of the doping concentration $\eta \sim 10 \%$, 
which gives  an upper bound on 
$\alpha \sim {\cal O}[10] {\em meV}$. In the absence of an external magnetic 
field, this coupling is relatively strong for $N=1$ four-component spinors 
to produce a superconducting gap (\ref{flavour}) of order $0.15$ meV. 
This gap occurs at the nodes of the $d$-wave gap and the absence of a local 
order parameter~\cite{dor}, as reviewed in section 3, preserves 
the $d$-wave superconducting state. With a small gap of this size, the finite 
temperature analysis (\ref{gapT}) indicates a transition temperature
to a non-superconducting state of $T_c \simeq {\cal O}(0.1 K)$, so this gap 
(at the nodes of the $d$-wave gap) would have disappeared at much lower
temperatures, than the standard $T_c \simeq 100 K$ of the high-$T_c$ 
cuprates. However, as mentioned previously, the effective non-trivial 
running~\cite{amelino} 
number entering the gap formula (\ref{flavour})
may lead to a siginificant enhancement of the superconducting gap, 
and therefore of the respective critical temperature. 

The phenomenon (\ref{fit}), 
discussed above, implies an enhancement 
of the 
gap by external magnetic fields in (surface) 
regions of the superconductor,
where the external magnetic field can penetrate due to the Meissner effect.
For instance, for magnetic fields of order ${\cal O}[10]$ Tesla the 
gap may be enhanced up 
to: $m_B(B=10~{\rm Tesla}) \simeq 0.35\alpha
\simeq  3.5 meV$, 
which is smaller than the classical result (\ref{holongap}),
thereby justifying, to some extent, 
the ladder approximation, and the smallness of the 
quantum 
corrections assumed in the above calculations. 
This leads to a transition 
temperature (\ref{empiricalth2}) of ${\cal O}(30)$ K.

It is understood  that such numbers should only be viewed as indicative,  
since technically the above analysis is valid only 
for magnetic fields $B \rightarrow \infty$. 
More elaborate, analytic or lattice, treatments
are therefore necessary before definite conclusions are reached. 
Moreover, due to the Meissner effect the magnetic 
field is not uniformly distributed in space, something which according to the 
above analysis might lead to spatial anisotropies of the proposed gap $m_B$. 
The above estimates, however,   
appear to be reasonable enough to encourage further studies.

\begin{centering}
\begin{figure}[htb]
\epsfxsize=5in
\centerline{\epsffile{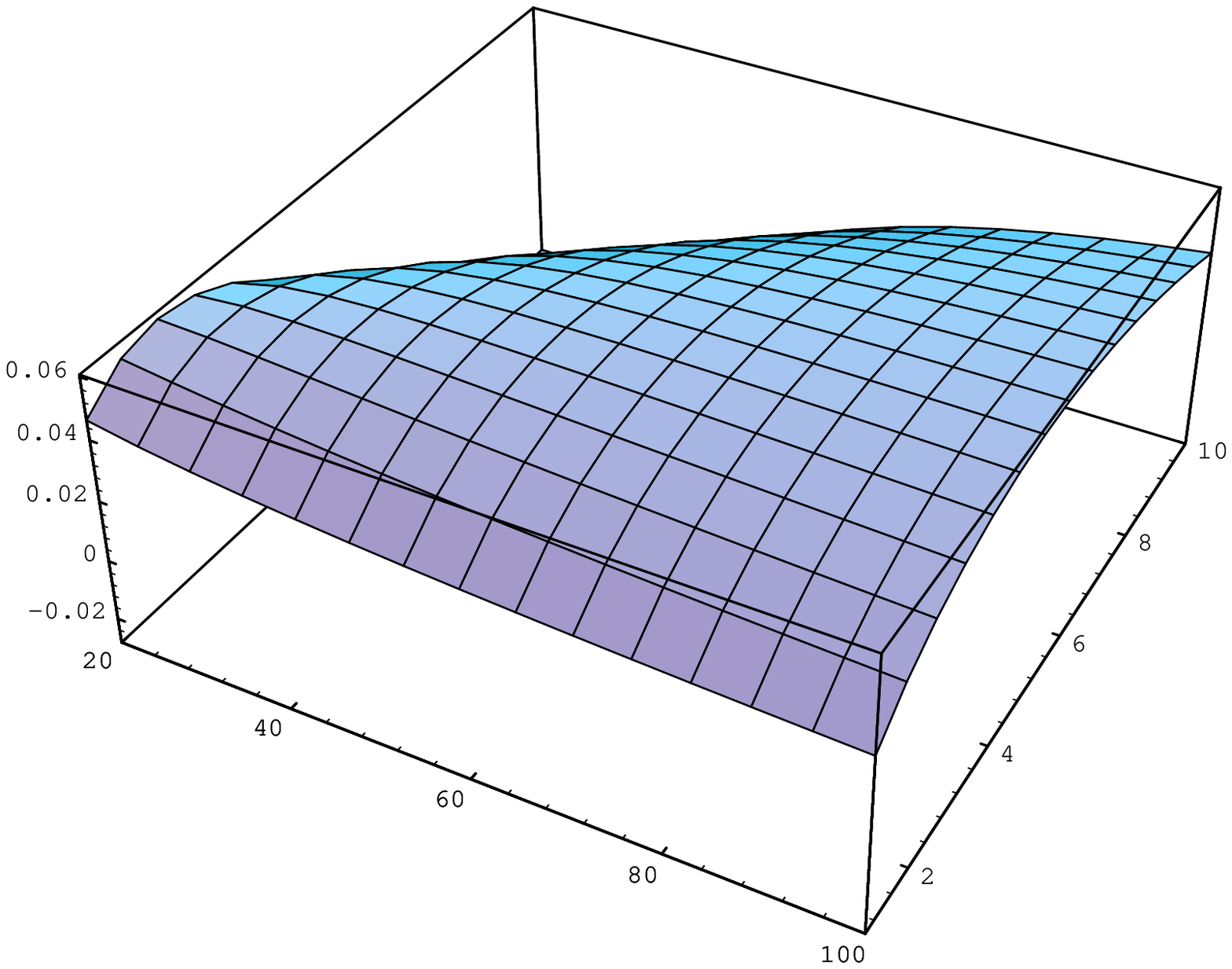}}
\vspace{0.01cm}
\caption{{\it 3D Plot of 
$\frac{\alpha}{2\pi}{\rm ln}[\sqrt{2x}/\alpha]/\sqrt{2x}$.
versus $(x,\alpha )$.  
The front horizontal axis denotes  $x$. 
For the value of $\alpha = 5$ the ratio remains almost fixed 
at $\sim 0.06$ (variation less than $5\%$ ), 
over the range of $x \sim 40 - 100$. }}
\label{fig6}
\end{figure}
\end{centering}

Let us now compare 
the above-described situation with the pertinent 
experimental results of ref. \cite{ong}, so as to get a 
better feeling on the physical meaning 
of the above results.   
In that work the authors find that, 
for strong enough magnetic fields 
of ${\cal O}[1]-{\cal O}[10]$ Tesla, there are 
plateaux in the 
thermal conductivity of quasiparticle excitations 
about the $d$-wave state (in particular about the nodes 
of the d-wave superconducting gap). Such plateux are  interpeted 
as an indication of the opening of a new gap, induced by the magnetic fields 
at the nodes. The authors find  that the plateaux disappear at a critical
temperature that depends on the magnetic field intensity, and in particular 
they  
report the empirical relation: 
\be
    T^{exp}_c \propto \sqrt{|eB|} 
\label{empirical}
\ee
for the dependence of the observed critical temperature 
with the external magnetic field strength.
Such a dependence 
is not obtained in our treatment  
of three-dimensional relativistic 
fermions in a strong external magnetic field.
In our analysis above, the result (\ref{empiricalth})  
for the critical 
temperature 
in the model of \cite{fm}
was mainly 
due to the {\it quantum dynamics } of the 
$U_S(1)$ gauge field, and 
exhibits only  
a logarithmic dependence  
on the magnetic field strength.
Notice that 
in four space-time dimensions the analogous relation  
indeed exhibits a $\sqrt{B}$ dependence~\cite{miransky}
\be
      T_c^{(4)} \sim \sqrt{|eB|}{\rm exp}[-\sqrt{\frac{\pi}{\alpha^{(4)}}}] 
\sim n'' m_B^{(4)}(T=0) 
\label{fourdim2}
\ee
where $n'' \sim {\cal O}[1]$, $m_B(T=0)$ is the zero-temperature
induced gap in (3+1) dimensions, and 
$\alpha^{(4)}$ is the four-dimensional Q.E.D. fine-structure constant,
which, unlike three dimensions, is dimensionless.

{}From this, one would be tempted to conclude that 
the charged quasiparticle excitations 
that play a role in the phenomenon of \cite{ong} 
might be genuinely four-dimensional. However, 
due to the small value of the $Q.E.D.$ 
fine-structure constant $\alpha ^{(4)} =1/137$ 
in four dimensions, the result (\ref{fourdim2}) would lead to 
unobservably small $T_c$. Does this, therefore, mean, that our model 
of relativistic three-dimensional fermions 
fails completely the experimental results of \cite{ong}?

We shall argue below that the answer to this question is no, at least 
for the time being. Indeed, by carefully looking at the 
experimental data, we observe~\footnote{We thank B.C. Georgalas 
for a discussion on this point.} that 
for the range of the magnetic field intensitites 
in the experiments of ref. \cite{ong} the available data 
{\it cannot exclude } the presence 
of Dirac fermions 
(holons) about 
the nodes of the d-wave gap, predicted by the model
of \cite{fm}, which lead to a condensate (\ref{fit}) 
with a logarithmic dependence
on the magnetic field. 
This is mainly due to the fact that in this model
one has an extra dimensionful scale $\alpha$, 
which may be viewed as an adjustable 
phenomenological parameter at this stage. 
For the regime of parameters 
relevant to the experimental data of \cite{ong},
in particular for magnetic fields of order $B=1 - 10~{\rm Tesla}$, 
the functions 
$\frac{\alpha}{2\pi}{\rm ln}(\sqrt{2|eB|}/\alpha )$ and $\sqrt{2|eB|}$,
which determine the critical temperature in (\ref{empiricalth}) and 
(\ref{empirical}), respectively,  
exhibit 
{\it similar variations } with the magnetic field $B$, 
{\it provided that}  
$\alpha \sim 5 ~{\rm meV}$. 
Indeed, as 
can be 
easily deduced by plotting their ratio  
(figure 6),  for $\alpha = {\cal O}(5)$ meV and $\hbar v_F 11.55 \sqrt{eB}$ 
of order 
${\cal O}(\sqrt{40}) - {\cal O}(\sqrt{400})$ meV, 
i.e. $B={\cal O}(1)-{\cal O}(10)$ Tesla, 
one finds that $\frac{\alpha}{2\pi}{\rm ln}(\sqrt{2|eB|}/\alpha ) 
\simeq 0.06 \sqrt{ 2eB}$ 
(in units of $\hbar  v_F=1$ with $\alpha$, $\sqrt{|eB|}$,
expressed in meV), with deviation of order $5-10\%$. 
This means that, without significant error,  
one can 
express the critical temperature (\ref{empiricalth}) 
in the model of ref. \cite{fm} 
for the magnetically-induced condensate also in the form (\ref{empirical}),  
with 
\be 
T_c \simeq 0.9\hbar v_F \sqrt{|eB|} =0.45 \sqrt{|eB|} ~{\rm meV}, 
  \quad B={\cal O}(1)-{\cal O}(10) 
~{\rm Tesla}, \quad \alpha ={\cal O}(5)~{\rm meV} 
\label{empfit}
\ee
ranging from $5.4$ K for $B \sim 1$ Tesla, to $17.4$ K for $B \sim 10$ Tesla. 
Such values are in agreement with the experimental findings of \cite{ong} 
for the above range of the magnetic fields. 
Notice also that the value of $\alpha \sim 5$ meV, for which the above
analysis is valid, 
is in agreement with the generic 
features 
(\ref{connection}) of the model of ref. \cite{fm}, and also 
with the Schwinger-Dyson (preliminary) analysis made above,
as it is smaller than the magnetic length $\sqrt{2|eB|}$.   
Thus, one may fit most of the available experimental data 
on $T_c$ using our relation (\ref{empiricalth}).
We consider this as evidence in favour of the relevance 
of our gauge model to the physics of high-temperature superconductivity. 
It is the low-magnetic-field regime, where at present 
there are not many data, which 
may lead to decisive conclusions about the validity of our model. 
Note, however, that in the low-magnetic field  
case the above preliminary analysis is not valid,
and one has to resort to lattice simulations. This is in 
progress~\cite{future}. 

We next remark that, in our analysis 
above, the third dimension has been 
ignored completely. 
In realistic models,   
the layer structure of the materials could be 
mimicked by 
viewing the system periodic along the third spatial dimension,
with period equal to the interlayer distance, estimated to be of order 
$100$ Angstrom in the 
high-temperature cuprates. Formally, this would imply a Fourier
series expansion of the corresponding (four-dimensional) 
gap equations
along the third direction. This might yield (c.f. (\ref{fourdim2})) 
a $\sqrt{B}$ dependence
for the induced gap~\cite{miransky}, 
even for the quantum corrections to (\ref{holongap}), and hence 
one could recover (\ref{empirical}).   
However, any conclusions would be premature at this stage. 

It is clear from the above discussion that 
a complete analysis
of the three dimensional systems, both analytic (Schwinger-Dyson) 
and via lattice simulations, 
is required before any definite conclusions 
are drawn. Although there are encouraging preliminary results, 
however, 
many issues, like the behaviour of the system 
under weak magnetic fields, the connection with the zero-field case, 
the very existence of a critical field 
(phase transition), below which the gap is purely due to 
the $U_S(1)$ statistical interactions for strong coupling, etc., still remain 
unanswered. The study of such effects requires detailed analyses
and proper lattice simulations, which 
are in progress~\cite{future}. Moreover, more input from experiment,
such as data on the gap at the low-magnetic-field (or even zero-field ) 
phase, is required to provide sufficient 
tests of the predictions made by the model of ref. \cite{fm} in connection 
with  
the above issues. 

Nevertheless, 
what becomes an important lesson from the above, admittedly crude,  
analysis is the fact that in the 
above model for high-$T_c$ dynamics 
there is an {\it even} number of fermion flavours~\cite{fm,dor}, and 
hence the induced
self-consistent gap for those fermions, which opens up 
as a result of the influence 
of the strong external magnetic field, is {\it not necessarily parity 
violating}. A {\it parity conserving}, self-consistent, solution to the 
induced gap equation (\ref{empiricalth}) has been found. However,  
at this stage, we are not in a definite position to 
exclude the possibility 
that a strong magnetic field induces a parity and time-reversal 
violating mass gap. 
This is the scenario suggested in ref. \cite{laughlinT}, 
as a possible explanation 
of the experimental results of ref. \cite{ong}.  
The presence of the external magnetic field 
is definitely an {\it external} source, violating time-reversal symmetry (and 
parity), and it is known that such sources may induce 
parity violating condensates in certain cases, 
provided they are strong enough~\cite{ambjorn}. 
Therefore, although 
in the above scenario we have found no evidence for such a phenomenon,
however, a complete analysis at an effective potential level,
including parity-violating dynamical condensates, along the lines of ref. \cite{app,Vafa},
is still lacking. 
This will be the topic of 
future work~\cite{future}.

\paragraph{}
\noindent {\Large {\bf Acknowledgements}}
\paragraph{}
N.E.M would like to thank 
the organisers of the {\it 5th Chia Workshop 
on Common Trends in Condensed Matter and Particle Physics}, 
for the invitation, and for providing 
a thought-stimulating atmosphere during the meeting. 
He would also like to thank R.B. Laughlin, for explaining his 
ideas on a possible interpretation of the results of ref. \cite{ong}
in terms of a magnetically-induced transition to a
Parity- and Time-Reversal- violating 
ground state. 
The authors   
would like to thank I. Aitchison, B. Georgalas, 
G. Koutsoumbas,
G. Semenoff and P. Wiegmann for discussions,  
and 
the CERN Theory 
division for hospitality during the last stages of this work. 
K.F. wishes to acknowledge partial financial support 
from PENED 95 Program, No. 1170, of the Greek General Secretariat
of Research and Technology.

\newpage

{\bf Note Added} 

\begin{itemize} 

{\bf \item{A}}. We would like to 
point out that an independent study of $QED_3$ 
in external magnetic fields has been made by Shpagin~\cite{shpagin},
who has performed 
a Schwinger-Dyson zero-temperature analysis including loops. 
For large magnetic fields, our analysis, based on dimensional reduction
of a ladder four-dimensional theory,  
agrees essentially with his,
as far as the logarithmic dependence of the dynamical mass $m_B$ 
on the magnetic field is concerned. From his precise analysis, 
however,one can also get information on subleading  corrections to $m_B$
of the form 
\be
-\alpha \left(\frac{1}{ 1 + {\cal O}(\alpha/\sqrt{eB})}
\left(1-\frac{1}{{\cal O}[{\rm ln}(\sqrt{eB}/\alpha)]}\right)
{\rm ln}[{\rm ln}(\sqrt{eB}/\alpha)] \right) 
\nn 
\ee
Notice also that, as shown 
in ref. \cite{shpagin}, the photon vacuum polarization 
is suppressed for strong magnetic fields
by terms $\alpha/\sqrt{eB}$, thereby justifying the ladder approximation    
used in our qualitative analysis in this work. 
We thank V. Gusynin and V. Miransky for pointing out this reference to us, and 
for a useful discussion.

{\bf \item{B}}. An important aspect of 
the superconducting 
model of ref. \cite{fm}, described above, was the appearance 
of a {\it second superconducting phase}, due to the opening of a gap 
at the nodes of a $d$-wave gap. This gap was caused by 
two species of relativistic fermions, 
which acquired dynamically a {\it parity-conserving} mass (\ref{flavour}),
disappearing at a temperature (\ref{gapT}). 
Due to the relativistic nature of the problem this gap is of $s$-wave type. 
As discussed in section 5, 
for values of the dimensionful coupling constant $\alpha \sim {\cal O}(10)$
meV, consistent with the microscopic scenario~\cite{fm} 
for magnetic superconductivity of the model (\ref{su2action}), 
one obtains a transition temperature of order of a few hundreds of mK, 
which leads to a second superconducting phase. This phase appears to be  
in addition to the 
d-wave high-$T_c$ superconductivity, with $T_c ={\cal O}(100)$ K,  
which is due to 
the bulk of the fermi surface for holons in the model.

This result, which, from the point of view of the present work, 
pertains to the 
zero (or weak) external magnetic field case, may be related to the 
recent experimental findings of ref. \cite{recent}. These authors 
report 
on a low-temperature superconducting phase
in Ni-doped BiSrCaCuO high-temperature cuprate, 
with $T_c \simeq 200$ mK. Note that  
the standard (d-wave) high-$T_c$ superconducting phase for this material 
has $T_c \simeq 77$ K. This second superconducting phase is discovered
by looking at the thermal 
conductivity of quasiparticles, 
as
in the experiment
of ref. \cite{ong}~\footnote{For comparison, 
we note that the experiment of ref. \cite{ong} used 
samples of BiSrCaCuO, i.e. their case corresponds to the zero-Ni-doping case 
of \cite{recent}.}, and 
was related in \cite{recent} to the opening of a gap at the nodes of the 
$d$-wave gap. The fact that the transition is observed after doping with 
Ni magnetic impurities may be suggestive of the fact that 
a {\it small} magnetic field is responsible for triggering the phenomenon,
which from our point of view here 
would be similar to the magnetic catalysis phenomenon 
of chiral symmetry breaking,  discussed above. 
In the paper of ref. \cite{recent} a suggestion was made, following ref. 
\cite{laughlinT}, that this transition signals time-reversal
breaking by the condensate, which, under the influence of the magnetic field, 
changes its state from 
the one described by an order parameter $d$ to that described 
by a complex Time-reversal violating order parameter $d + i d$. 
An experimental signature of such a 
state would be edge currents~\cite{laughlinT, volovik}.

However, as we mentioned in the text,  
time-reversal violation may not be necessary 
to explain the phenomenon. 
Indeed, 
the presence of magnetic impurities 
in Ni-doped BiSrCaCuO may admit an alternative interpretation, 
which matches our theoretical analysis in section 2, 
namely, such impurities 
provide an environment that results in 
a strong magnetic pairing attraction among the 
(relativistic) fermion (holon) excitations
at the nodes~\cite{fm,dor}. 
This is supported by the fact that the order of magnitude 
of the  transition (\ref{gapT}) 
agrees with the order of magnitude  
observed in ref. \cite{recent}. 
Of course, above a critical Ni-doping concentration, 
supoerconductivity will be destroyed, a feature which also seems 
to characterise the model fo ref. \cite{fm}. 

In either case,    
the 
result 
is the opening of a small $s$-wave Kosterlitz-
Thouless (KT) gap at the nodes
of the $d$-wave gap. 
A 
strong  external magnetic field, then,  
may enhance this gap, as explained above, without inducing 
parity or time-reversal violation. 
Moreover, the smooth character of the 
specific-heat curves, measured
in the experiment of \cite{recent}
in connection with the order of the second 
phase transition,
calls for comparison with 
the Kosterlitz-Thouless nature of the superconducting gap  
in the model of refs. \cite{fm,dor}. 
It will be interesting to explore further the r\^ole of relativistic fermion
superconductivity, and the effect of magnetic fields in the context
of our microscopic statistical models~\cite{fm}, 
by computing the thermal 
conductivity of quasiparticles, and discussing the r\^ole of 
magnetic impurities, vortices etc  
in such a context.
These will be left for future work. 

\end{itemize}

\end{document}